\documentclass[aps,prl,twocolumn,superscriptaddress]{revtex4-1}
\usepackage{bm} \usepackage{graphicx} \usepackage{amsmath}
\usepackage{amsthm,stmaryrd}
\usepackage{amssymb} \usepackage{epstopdf} \usepackage{times,txfonts}
\usepackage{color}
\usepackage{ytableau}
\usepackage{verbatim} 
\usepackage{soul} 
\usepackage{subfigure}
\usepackage{easybmat}
\usepackage{multirow}
\usepackage{listings}
\usepackage{diagbox}
\usepackage{mathtools}
\usepackage{array}
\usepackage[bookmarks=false]{hyperref}
\hypersetup{colorlinks=true,citecolor=blue,linkcolor=blue,urlcolor=blue,pdfstartview=FitH,bookmarksopen=true}

\DeclareRobustCommand\openzero{\leavevmode\hbox{0\kern-.55em0}}
\mathchardef\minus="002D

\begin{document}

\title{
  $d+1$ Measurement Bases are Sufficient for Determining $d$-Dimensional Quantum States:\\
  Theory and Experiment
}

\author{ Tianqi Xiao }
\affiliation{School of Physics and State Key Laboratory of Optoelectronic Materials and Technologies, Sun Yat-sen University, Guangzhou 510000, China}

\author{ Yaxin Wang }
\affiliation{School of Physics and State Key Laboratory of Optoelectronic Materials and Technologies, Sun Yat-sen University, Guangzhou 510000, China}

\author{ Ying Xia }
\affiliation{School of Physics and State Key Laboratory of Optoelectronic Materials and Technologies, Sun Yat-sen University, Guangzhou 510000, China}

\author{ Zhihao Li }
\affiliation{School of Physics and State Key Laboratory of Optoelectronic Materials and Technologies, Sun Yat-sen University, Guangzhou 510000, China}

\author{ Xiaoqi Zhou }
\email{zhouxq8@mail.sysu.edu.cn}
\affiliation{School of Physics and State Key Laboratory of Optoelectronic Materials and Technologies, Sun Yat-sen University, Guangzhou 510000, China}
\affiliation{Hefei National Laboratory, University of Science and Technology of China, Hefei 230088, China}

\date{\today}
\begin{abstract}
A long-standing problem in quantum physics is to determine the minimal number of measurement bases required for the complete characterization of unknown quantum states, a question of particular relevance to high-dimensional quantum information processing. Here, we propose a quantum state tomography scheme that requires only $d+1$ projective measurement bases to fully reconstruct an arbitrary $d$-dimensional quantum state. As a proof-of-principle, we experimentally verified this scheme on a silicon photonic chip by reconstructing quantum states for $d=6$, in which a complete set of mutually unbiased bases does not exist. This approach offers new perspectives for quantum state characterization and measurement design, and holds promise for future applications in quantum information processing.
\end{abstract}

\maketitle

Quantum state tomography (QST) reconstructs the density matrix of an unknown quantum state by measuring many identical copies \cite{paris2004quantum,haffner2005scalable}. This technique is essential in quantum computing \cite{divincenzo1995quantum,o2007optical}, quantum communication \cite{bennett1993teleporting,bouwmeester1997experimental}, and quantum metrology \cite{giovannetti2006quantum,giovannetti2011advances}. For a $d$-dimensional state, the standard approach uses a number of measurement bases of order $d^2$ and reconstructs the density matrix from the expectation values of measurement operators \cite{james2001measurement,thew2002qudit,hioe1981n,bertlmann2008bloch,kimura2003bloch,mendavs2006classification}. Each measurement provides both an expectation value and a probability distribution, but existing methods do not fully use all the information available. Making fuller use of the probability data can reduce the number of measurement bases required. Based on this idea, recent studies have shown that only $2d-1$ projective measurement bases are sufficient for complete state reconstruction in any $d$-dimensional quantum system \cite{newton1968measurability,perlin2021spin,wang2024direct}.

Another prominent strategy for QST is based on the construction of mutually unbiased bases (MUBs) \cite{ivonovic1981geometrical,wootters1989optimal,zauner2011quantum,durt2010mutually,horodecki2022five,mcnulty2024mutually}. In a $d$-dimensional Hilbert space, at most $d+1$ MUBs can exist in theory. If a complete set of $d+1$ MUBs can be constructed for a given dimension, QST can be performed using only these bases. However, current research shows that a complete set of $d+1$ MUBs can be constructed only when $d$ is a power of a prime \cite{bandyopadhyay2002new,klappenecker2004constructions,archer2005there,brierley2009all,adamson2010improving,lima2011experimental,giovannini2013characterization}. In other dimensions, this construction remains an open problem; for example, for $d=6$, only three MUBs have been found to date~\cite{bengtsson2007mutually,grassl2004sic,jaming2009generalized,butterley2007numerical,brierley2009constructing,raynal2011mutually,goyeneche2013mutually}. As a result, MUB-based QST lacks universality and cannot be implemented in all dimensions. This raises a natural question: for any dimension $d$, can QST be achieved with fewer than $2d-1$ measurement bases?

In this Letter, we present a new QST scheme that requires only $d+1$ projective measurement bases to achieve full reconstruction of arbitrary $d$-dimensional quantum states. We further experimentally demonstrate this scheme by reconstructing six-dimensional single-photon states on a silicon photonic chip, achieving high-fidelity results in a dimension where a complete set of MUBs does not exist. Our scheme not only offers theoretical insight into the fundamental structure of quantum measurements, but also provides a new perspective for designing measurement protocols in high-dimensional quantum systems.

Before detailing our scheme, we first analyze the minimal number of measurement bases required for QST from the perspective of parameter counting. For a $d$-dimensional quantum state, the density matrix
\begin{equation}
\rho =\sum_{a,b=0}^{d-1}{g_{a,b}|a\rangle \langle b|},
\label{eq:rho}
\end{equation}
is Hermitian, normalized as $\mathrm{Tr}(\rho) = 1$, and contains $d^2-1$ independent real parameters. Measuring the state in each orthonormal basis yields $d$ probability values, but at most $d-1$ of them are independent. Therefore, to fully determine the density matrix, at least $d+1$ distinct measurement bases are required.

\begin{figure*}[!t]
    \centering
    \includegraphics[width=\textwidth]{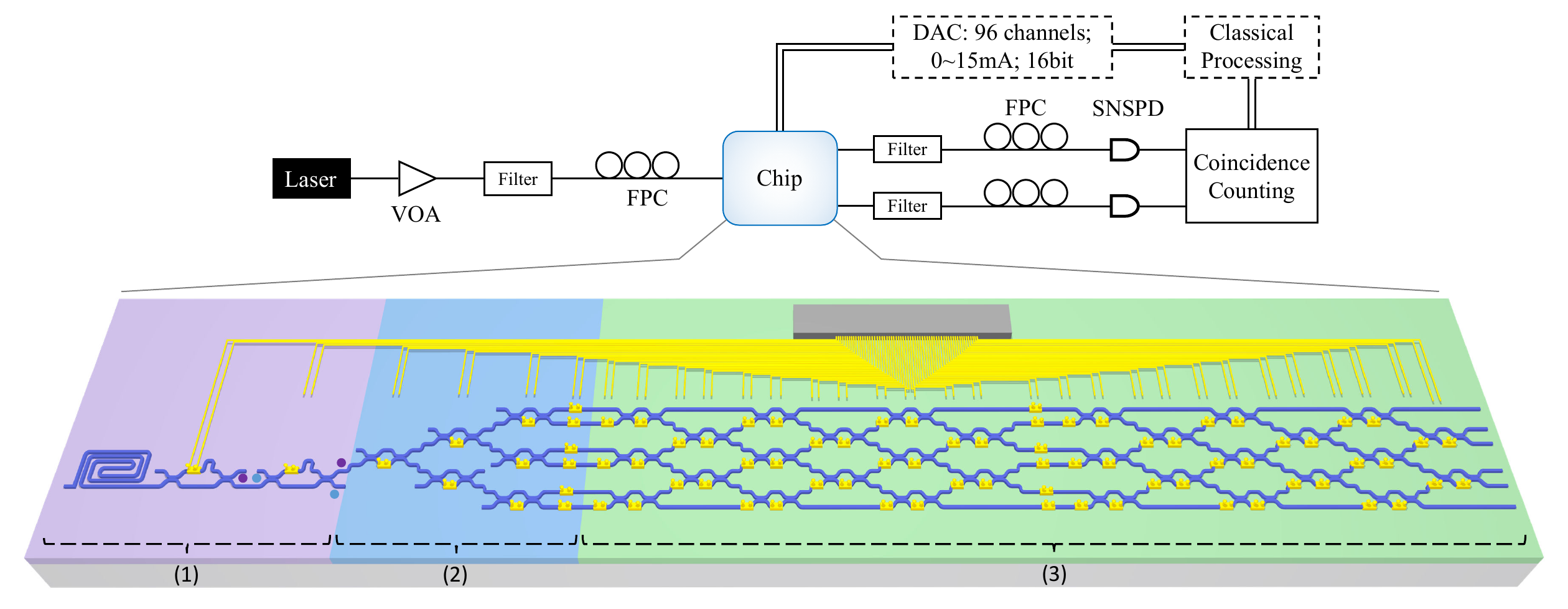}
    \caption{
    Schematic of the silicon quantum photonic chip and external setup. The system consists of three functional modules: (1) Single-photon generation via four-wave mixing in a silicon waveguide spiral; (2) Preparation of six-dimensional path-encoded quantum states using multiple Mach-Zehnder interferometers (MZIs); (3) Quantum state measurements in seven projective measurement bases, yielding seven sets of probability distributions.
    VOA, variable optical attenuation; FPC, fiber polarization controller; SNSPD, superconducting nanowire single-photon detector.
    }
    \label{chip}
\end{figure*}

We now present a QST scheme that applies to arbitrary dimension $d$, requiring only $d+1$ distinct projective measurement bases for complete state reconstruction, which matches the theoretical minimum. For any $d$-dimensional quantum state $\rho$, the $d+1$ measurement bases are chosen as follows. The first basis is the computational basis, $\{ |0\rangle, |1\rangle, \ldots, | d-1 \rangle \}$. The second is the Fourier basis,
\begin{equation}
\{\, |\psi_k^{(0)}\rangle = \frac{1}{\sqrt{d}} \sum_{l=0}^{d-1} \omega^{kl} |l\rangle \,\}_{k=0,1,\ldots,d-1},
\label{eq:psi}
\end{equation}
where $\omega = e^{2\pi i / d}$. The remaining $d-1$ measurement bases are generated by applying $d-1$ different diagonal unitary operators,
\begin{equation}
\begin{aligned}
R^{(j)} = \sum_{m=0}^{d-1} e^{i \theta_m^{(j)}} |m\rangle \langle m|,
\label{eq:R}
\end{aligned}
\end{equation}
with $j = 1, 2, \ldots, d-1$, to the Fourier basis. By defining $R^{(0)}$ as the identity operator, the second through the $(d+1)$-th measurement bases can be uniformly expressed as
\begin{equation}
\begin{aligned}
&\{\, |\psi_k^{(0)}\rangle = R^{(0)} |\psi_k^{(0)}\rangle \,\}_k, \\
&\{\, |\psi_k^{(1)}\rangle = R^{(1)} |\psi_k^{(0)}\rangle \,\}_k, \\
&\quad\quad\quad \ldots, \\
&\{\, |\psi_k^{(d-1)}\rangle = R^{(d-1)} |\psi_k^{(0)}\rangle \,\}_k.
\end{aligned}
\end{equation}
Measurements in these $d+1$ projective bases yield $d+1$ sets of probability distributions. The distribution from the first basis (the computational basis) is $\{ b_0, b_1, \ldots, b_{d-1} \}$, giving the probability for each outcome. The remaining $d$ sets, corresponding to the Fourier basis and its variants under diagonal unitary transformations, are
\begin{equation}
\begin{aligned}
&\{ p_0^{(0)}, p_1^{(0)}, \ldots, p_{d-1}^{(0)} \}, \\
&\{ p_0^{(1)}, p_1^{(1)}, \ldots, p_{d-1}^{(1)} \}, \\
&\{ p_0^{(2)}, p_1^{(2)}, \ldots, p_{d-1}^{(2)} \}, \\
&\quad\quad\quad \ldots, \\
&\{ p_0^{(d-1)}, p_1^{(d-1)}, \ldots, p_{d-1}^{(d-1)} \}.
\end{aligned}
\end{equation}

These $d+1$ probability distributions are sufficient to fully reconstruct the density matrix $\rho$. Specifically, the diagonal elements are directly determined by the probability distribution in the computational basis,
\begin{equation}
\begin{aligned}
g_{k,k} = b_k,
\label{eq:b}
\end{aligned}
\end{equation} 
where $k = 0, 1, \ldots, d-1$.
The off-diagonal elements are extracted from the remaining $d$ probability distributions, via
\begin{equation}
\begin{aligned}
p_k^{(j)} = \langle \psi_k^{(j)} | \rho | \psi_k^{(j)} \rangle =\langle \psi _{k}^{(0)}|R^{(j) \dagger}\rho R^{(j)}|\psi _{k}^{(0)}\rangle.
\label{eq:pjk}
\end{aligned}
\end{equation} 
By substituting Eqs.~(\ref{eq:rho})--(\ref{eq:R}) into this expression, one can obtain
\begin{equation}
\begin{aligned}
p_{k}^{(j)}= \frac{1}{d} + \frac{1}{d} \sum_{\substack{a,b=0 \\ a \ne b}}^{d-1} \omega^{-k(a-b)} e^{-i(\theta_a^{(j)} - \theta_b^{(j)})} g_{a,b}.
\label{eq:pjk2}
\end{aligned}
\end{equation}
Define $q_{k}^{(j)} = d p_{k}^{(j)} - 1$ so that
\begin{equation}
\begin{aligned}
q_{k}^{(j)} = \sum_{\substack{a,b=0 \\ a \ne b}}^{d-1} \omega^{-k(a-b)} e^{-i(\theta_a^{(j)} - \theta_b^{(j)})} g_{a,b},
\label{eq:pjk3}
\end{aligned}
\end{equation} 
where $j = 0, 1, \ldots, d-1$ and $k = 0, 1, \ldots, d-2$, with $k = d-1$ excluded since at most $d-1$ probabilities are independent for each measurement basis. Collecting all $q_k^{(j)}$ and $g_{a,b}$ ($a \ne b$) into $d(d-1)$-dimensional column vectors $\bm{q}$ and $\bm{g}$, Eq.~\eqref{eq:pjk3} can be written compactly as
\begin{equation}
\bm{q} = T \bm{g},
\end{equation}
where $T$ is a fixed $d(d-1) \times d(d-1)$ coefficient matrix with elements given by the coefficients in Eq.~\eqref{eq:pjk3}. It can be shown that $T$ is invertible if the phases ${\theta_m^{(j)}}$ satisfy the following two conditions:
\begin{equation}
\begin{aligned}
\theta_t^{(1)} - \theta_{(t+c)\bmod d}^{(1)} &\neq \theta_{t’}^{(1)} - \theta_{(t’+c)\bmod d}^{(1)} + 2\pi n, \\
\theta_m^{(j)} &= j\theta_m^{(1)},
\label{eq:cond}
\end{aligned}
\end{equation}
for $t \neq t’$, $t, t’ = 0,1,\ldots,d-1$, $c = 1,2,\ldots,d-1$, $n=0,\pm 1,\pm 2,\ldots $, and $j, m = 0,1,\ldots,d-1$ (see Supplemental Material).
The off-diagonal elements, collected in $\bm{g}$, are then uniquely determined by
\begin{equation}
\begin{aligned}
\bm{g}=T^{-1}\bm{q},
\label{eq:g}
\end{aligned}
\end{equation} 
thus enabling full reconstruction of the density matrix $\rho$.


\begin{figure*}[!t]
    \centering
    \includegraphics[width=\textwidth]{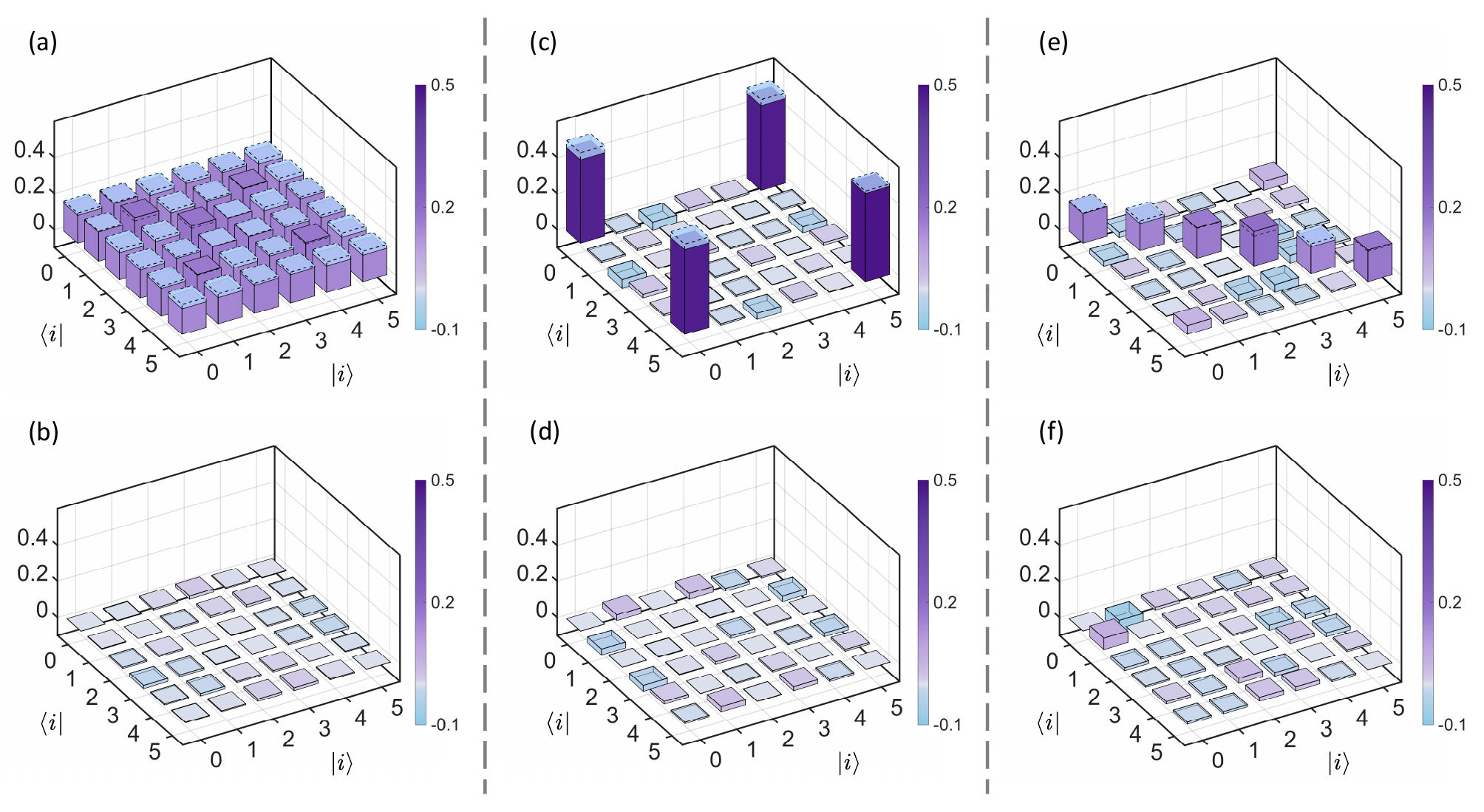}
    \caption{
    Experimentally reconstructed density matrices. (a), (c), and (e) show the real parts, while (b), (d), and (f) show the imaginary parts of the reconstructed density matrices for the three six-dimensional quantum states. Dashed lines indicate the ideal values.
    }
    \label{tomo}
\end{figure*}

To experimentally verify this QST scheme, we developed a programmable quantum optical platform based on a silicon photonic chip. As shown in Fig.~\ref{chip}, this platform integrates all essential components, from single-photon generation to quantum state measurement.
(1) Single-photon generation.
A pulsed laser operating at 80 MHz with a central wavelength of 1550.3 nm is injected into the silicon waveguide spiral on the chip, where photon pairs are produced through the process of four-wave mixing. The pump laser is first removed by an asymmetric Mach-Zehnder interferometer (AMZI). Subsequently, a second AMZI separates the generated photons, with the idler photon (wavelength 1553.5 nm) used as a trigger and the signal photon (wavelength 1547.1 nm) routed to the downstream quantum circuits for further processing.
(2) Preparation of six-dimensional quantum states.
The signal photon then passes through an interferometric network composed of six Mach-Zehnder interferometers (MZIs), where six heaters allow arbitrary amplitude modulation for each path. Each path is also equipped with an independent heater for phase control, enabling the preparation of any pure six-dimensional quantum state. By preparing up to six such pure states and mixing them with appropriate probabilities, any six-dimensional mixed state can be realized.
(3) Quantum state measurement and data acquisition.
The six-dimensional quantum state then passes through a reconfigurable six-dimensional linear optical network on the chip, consisting of multiple MZIs and heaters. By adjusting the heaters, arbitrary six-dimensional unitary transformations can be realized \cite{clements2016optimal,li2023experimental}. In our experiment, seven distinct unitaries are implemented using a variational approach \cite{peruzzo2014variational,xue2022variational}, corresponding to seven projective measurement bases. The first measurement basis is the computational basis, and the other six are the Fourier basis and its variants under diagonal unitary transformations, where the diagonal phases are set to $\theta_m^{(j)} = 0.5671\, j m^2$, with $j = 1, 2, \ldots, 5$ and $m = 0, 1, \ldots, 5$ \cite{author1}. After the signal photon passes through the unitary transformation network, both the signal and idler photons are coupled out of the chip via a fiber array and detected by superconducting nanowire single-photon detectors (SNSPDs), from which the probability distributions for each measurement basis are extracted. Using these distributions, the full density matrix of any six-dimensional quantum state can be reconstructed by applying the theoretical scheme described above together with maximum likelihood estimation \cite{james2001measurement,thew2002qudit}.

In our experiment, we selected the following two six-dimensional pure states and one six-dimensional mixed state for QST:
\begin{gather}
\frac{1}{\sqrt{6}}(\, |0\rangle + |1\rangle + \ldots + |5\rangle \,), \notag\\
\frac{1}{\sqrt{2}}(\, |0\rangle + |5\rangle \,), \\
\frac{1}{6}(\, |0\rangle \langle 0| + |1\rangle \langle 1| + \ldots + |5\rangle \langle 5| \,). \notag
\end{gather}
Using our scheme, we experimentally reconstructed the density matrices of all three states, as shown in Fig.~\ref{tomo}. The fidelities between the reconstructed density matrices $\rho’$ and the ideal states $\rho$ are calculated as
\begin{equation}
\begin{aligned}
F=tr\sqrt{\rho ^{1/2}\rho' \rho ^{1/2}},
\end{aligned}
\end{equation}
yielding values of $0.9654 \pm 0.0067$, $0.9698 \pm 0.0042$, and $0.9761 \pm 0.0033$, respectively.

In summary, we have theoretically proposed and experimentally demonstrated a QST scheme that reconstructs $d$-dimensional quantum states using only $d+1$ projective measurement bases, matching the theoretical minimum. We implemented six-dimensional quantum state tomography on a silicon photonic chip, achieving fidelities above $0.96$, in close agreement with theoretical predictions. Our results verify the effectiveness of this scheme even in dimensions where a complete set of MUBs cannot be constructed, demonstrating its practical feasibility. This approach not only sheds new light on the fundamental structure of quantum measurements, but also provides new perspectives for quantum state characterization and measurement design, and may find applications in future quantum information processing.

\begin{acknowledgements} 
This work was supported by the National Key research and development Program of China (No.2021YFA1400800), the National Natural Science Foundation of China (Grant No.61974168). X.-Q. Zhou acknowledges support from the Innovation Program for Quantum Science and Technology (Grant No.2021ZD0300702).
\end{acknowledgements}

\end{document}


\title{
Supplemental Material for ‘‘$d+1$ Measurement Bases Are Sufficient for Determining\\ $d$‑Dimensional Quantum States: Theory and Experiment’’
}

\author{ Tianqi Xiao }
\affiliation{School of Physics and State Key Laboratory of Optoelectronic Materials and Technologies, Sun Yat-sen University, Guangzhou 510000, China}

\author{ Yaxin Wang }
\affiliation{School of Physics and State Key Laboratory of Optoelectronic Materials and Technologies, Sun Yat-sen University, Guangzhou 510000, China}

\author{ Ying Xia }
\affiliation{School of Physics and State Key Laboratory of Optoelectronic Materials and Technologies, Sun Yat-sen University, Guangzhou 510000, China}

\author{ Zhihao Li }
\affiliation{School of Physics and State Key Laboratory of Optoelectronic Materials and Technologies, Sun Yat-sen University, Guangzhou 510000, China}

\author{ Xiaoqi Zhou }
\email{zhouxq8@mail.sysu.edu.cn}
\affiliation{School of Physics and State Key Laboratory of Optoelectronic Materials and Technologies, Sun Yat-sen University, Guangzhou 510000, China}
\affiliation{Hefei National Laboratory, University of Science and Technology of China, Hefei 230088, China}

\date{\today}

\maketitle

\section*{Invertibility Analysis of the Reconstruction Matrix $T$}

In the main text, we have obtained the following equation:
\begin{equation}
\bm{q} = T \bm{g}.
\end{equation}
Here, both $\bm{q}$ and $\bm{g}$ are column vectors of dimension $d(d-1)$. 
Each entry of $\bm{g}$ corresponds to a nondiagonal element $g_{a,b}$ ($a \ne b$) of the density matrix $\rho$, while each entry of $\bm{q}$ is calculated from the measured probability distributions $p_k^{(j)}$ as $q_k^{(j)} = d p_k^{(j)} - 1$ (where $j = 0, 1, \ldots, d-1$ and $k = 0, 1, \ldots, d-2$).
The order of $\bm{q}$ is as follows: for each $j = 0, 1, \ldots, d-1$, all $q_k^{(j)}$ with $k = 0, 1, \ldots, d-2$ are arranged in sequence, i.e.,
\begin{equation}
\bm{q} = \left[ q_0^{(0)},\ q_1^{(0)},\ \ldots,\ q_{d-2}^{(0)},\ 
q_0^{(1)},\ q_1^{(1)},\ \ldots,\ q_{d-2}^{(1)},\ \ldots,\ 
q_0^{(d-1)},\ q_1^{(d-1)},\ \ldots,\ q_{d-2}^{(d-1)} \right]^{\mathsf{T}}.
\end{equation}
The order of $\bm{g}$ is as follows: first, all $g_{a,b}$ with $(b-a)\bmod d=1$ are listed, then those with $(b-a)\bmod d=2$, and so on, up to $(b-a)\bmod d=d-1$. 
Explicitly,
\begin{equation}
\bm{g} = \left[
g_{0,1},\ g_{1,2},\ \ldots,\ g_{d-2,d-1},\ g_{d-1,0},\
g_{0,2},\ g_{1,3},\ \ldots,\ g_{d-2,0},\ g_{d-1,1},\
\ldots,\
g_{0,d-1},\ g_{1,0},\ \ldots,\ g_{d-2,d-3},\ g_{d-1,d-2}
\right]^{\mathsf{T}}.
\end{equation}
$T$ is a $d(d-1) \times d(d-1)$ matrix whose elements are defined as
\begin{equation}
T_{(d-1)j + k,\, d(c-1) + a} = \omega^{kc}\, e^{-i(\theta_{a}^{(j)} - \theta_{(a+c) \bmod d}^{(j)})},
\end{equation}
where $j = 0, 1, \ldots, d-1$, $k = 0, 1, \ldots, d-2$, $c = 1, 2, \ldots, d-1$, $a = 0, 1, \ldots, d-1$. 
The overall structure of $T$ can be written as
\begin{equation}
T =
\left[
\begin{array}{ccccccccc}
1 & \cdots & 1 & & \cdots & & 1 & \cdots & 1 \\
\vdots & \ddots & \vdots & & \cdots & & \vdots & \ddots & \vdots \\
\omega^{d-2} & \cdots & \omega^{d-2} & & \cdots & & \omega^{(d-2)(d-1)} & \cdots & \omega^{(d-2)(d-1)} \\
& & & & & & & & \\
\vdots & \vdots & \vdots & & \ddots & & \vdots & \vdots & \vdots \\
& & & & & & & & \\
e^{-i(\theta_{0}^{(d-1)} - \theta_{1}^{(d-1)})} & \cdots & e^{-i(\theta_{d-1}^{(d-1)} - \theta_{0}^{(d-1)})} & & \cdots & & e^{-i(\theta_{0}^{(d-1)} - \theta_{d-1}^{(d-1)})} & \cdots & e^{-i(\theta_{d-1}^{(d-1)} - \theta_{d-2}^{(d-1)})} \\
\vdots & \ddots & \vdots & & \cdots & & \vdots & \ddots & \vdots \\
\omega^{d-2} e^{-i(\theta_{0}^{(d-1)} - \theta_{1}^{(d-1)})} & \cdots & \omega^{d-2} e^{-i(\theta_{d-1}^{(d-1)} - \theta_{0}^{(d-1)})} & & \cdots & & \omega^{(d-2)(d-1)} e^{-i(\theta_{0}^{(d-1)} - \theta_{d-1}^{(d-1)})} & \cdots & \omega^{(d-2)(d-1)} e^{-i(\theta_{d-1}^{(d-1)} - \theta_{d-2}^{(d-1)})}
\end{array}
\right].
\end{equation}

To analyze the invertibility of $T$, we left-multiply $T$ by two fixed full-rank matrices, $B_1^{-1}$ and $B_2^{-1}$, transforming it into a block-diagonal matrix $B_3$:
\begin{equation}
B_2^{-1} B_1^{-1} T = B_3.
\end{equation}
Here, $B_1$ is a $d(d-1) \times d(d-1)$ block-diagonal matrix,
\begin{equation}
\begin{aligned}
B_1=\left[ \begin{matrix}{}	M&		&		&		&		\\	&		M&		&		&		\\	&		&		M&		&		\\	&		&		&		\ddots&		\\	&		&		&		&		M\\\end{matrix} \right] ,
\end{aligned}
\end{equation}
where $M$ is a $(d-1) \times (d-1)$ matrix with entries $M_{k,c} = \omega^{kc}$.
$B_2$ is a $d(d-1) \times d(d-1)$ sparse matrix, defined as
\begin{equation}
\left( B_2 \right) _{\left( d-1 \right) j+k,\, d\left( c-1 \right) +a}=\delta _{j,a}\, \delta _{k,c-1}.
\end{equation}
$B_3$ is a $d(d-1) \times d(d-1)$ block-diagonal matrix,
\begin{equation}
\begin{aligned}
B_3=\left[ \begin{matrix}	N^{\left( 1 \right)}&		&		&		\\	&		N^{\left( 2 \right)}&		&		\\	&		&		\ddots&		\\	&		&		&		N^{\left( d-1 \right)}\\\end{matrix} \right] ,
\end{aligned}
\end{equation}
where each $N^{(c)}$ is a $d\times d$ matrix with entries $N_{j,a}^{\left( c \right)}=e^{-i( \theta _{a}^{\left( j \right)}-\theta _{\left( a+c \right) \bmod d}^{\left( j \right)} )}$. 
The explicit form of $N^{(c)}$ is given by
\begin{equation}
\begin{aligned}
N^{\left( c \right)}=\left[ \begin{matrix}	1&		1&		\cdots&		1\\
e^{-i( \theta _{0}^{\left( 1 \right)}-\theta _{c}^{\left( 1 \right)} )}&		e^{-i( \theta _{1}^{\left( 1 \right)}-\theta _{\left( 1+c \right) \bmod d}^{\left( 1 \right)} )}&		\cdots&		e^{-i( \theta _{d-1}^{\left( 1 \right)}-\theta _{\left( d-1+c \right) \bmod d}^{\left( 1 \right)} )}\\
\vdots&		\vdots&		\ddots&		\vdots\\
e^{-i( \theta _{0}^{\left( d-1 \right)}-\theta _{c}^{\left( d-1 \right)} )}&		e^{-i( \theta _{1}^{\left( d-1 \right)}-\theta _{\left( 1+c \right) \bmod d}^{\left( d-1 \right)} )}&		\cdots&		e^{-i( \theta _{d-1}^{\left( d-1 \right)}-\theta _{\left( d-1+c \right) \bmod d}^{\left( d-1 \right)} )}\\\end{matrix} \right] .
\label{eq:Nc}
\end{aligned}
\end{equation}
Since both $B_1$ and $B_2$ are full-rank, this linear transformation does not affect the invertibility of $T$. 
Therefore, $T$ is invertible if and only if $B_3$ is invertible.

When the phases ${\theta_m^{(j)}}$ satisfy
\begin{equation}
\theta_m^{(j)} = j\theta_m^{(1)},
\label{eq:condition1}
\end{equation}
each $N^{(c)}$ reduces to a Vandermonde matrix, with the explicit form
\begin{equation}
\begin{aligned}
N^{\left( c \right)}=\left[ \begin{matrix}	1&		1&		\cdots&		1\\
e^{-i( \theta _{0}^{\left( 1 \right)}-\theta _{c}^{\left( 1 \right)} )}&		e^{-i( \theta _{1}^{\left( 1 \right)}-\theta _{\left( 1+c \right) \bmod d}^{\left( 1 \right)} )}&		\cdots&		e^{-i( \theta _{d-1}^{\left( 1 \right)}-\theta _{\left( d-1+c \right) \bmod d}^{\left( 1 \right)} )}\\
\vdots&		\vdots&		\ddots&		\vdots\\
( e^{-i( \theta _{0}^{\left( 1 \right)}-\theta _{c}^{\left( 1 \right)} )} ) ^{d-1}&		( e^{-i( \theta _{1}^{\left( 1 \right)}-\theta _{\left( 1+c \right) \bmod d}^{\left( 1 \right)} )} ) ^{d-1}&		\cdots&		( e^{-i( \theta _{d-1}^{\left( 1 \right)}-\theta _{\left( d-1+c \right) \bmod d}^{\left( 1 \right)} )} ) ^{d-1}\\\end{matrix} \right] ,
\end{aligned}
\end{equation}
whose determinant is
\begin{equation}
\det ( N^{(c)} ) = \prod_{0\leq t’ < t \leq d-1} \left( e^{-i(\theta_t^{(1)} - \theta_{(t+c)\bmod d}^{(1)})} - e^{-i(\theta_{t’}^{(1)} - \theta_{(t’+c)\bmod d}^{(1)})} \right).
\end{equation}
If the diagonal phases further satisfy
\begin{equation}
\theta_t^{(1)} - \theta_{(t+c)\bmod d}^{(1)} \neq \theta_{t’}^{(1)} - \theta_{(t’+c)\bmod d}^{(1)} + 2\pi n,
\label{eq:condition2}
\end{equation}
for $t \neq t’$, $t, t’ = 0,1,\ldots,d-1$, $c = 1,2,\ldots,d-1$, $n=0,\pm 1,\pm 2,\ldots $, then all $N^{(c)}$ are invertible.
Since $B_3$ is block-diagonal with blocks $N^{(c)}$, the invertibility of all $N^{(c)}$ ensures the invertibility of $B_3$, and thus of $T$. 
Therefore, we have shown that $T$ is invertible whenever the phases ${\theta_m^{(j)}}$ simultaneously satisfy Eqs. \eqref{eq:condition1} and \eqref{eq:condition2}.